\documentclass{article}
\usepackage{microtype}
\usepackage{graphicx}
\usepackage{subcaption}
\usepackage{booktabs}
\usepackage{hyperref}
\usepackage{amsmath}
\usepackage{amssymb}

\usepackage[accepted]{icml2026}
\providecommand{\Cref}[1]{\ref{#1}}
\providecommand{\cref}[1]{\ref{#1}}
\newcommand{\Tabref}[1]{Table~\ref{#1}}

\newcommand{\Secref}[1]{Section~\ref{#1}}

\usepackage{xcolor}

\icmltitlerunning{Voting Protocols as Coordination Mechanisms for Role-Constrained Multi-Agent Tutoring Systems}

\begin{document}

\twocolumn[
  \icmltitle{Voting Protocols as Coordination Mechanisms \\ for Role-Constrained Multi-Agent Tutoring Systems}

  \icmlsetsymbol{equal}{*}
  \begin{icmlauthorlist}
  \icmlauthor{Eric S. Qiu}{equal,cornell}
  \icmlauthor{Joyce Gill}{equal,stanford}
  \end{icmlauthorlist}
  \icmlaffiliation{cornell}{Cornell University, Ithaca, New York, USA}
  \icmlaffiliation{stanford}{Stanford University, Stanford, California, USA}
  \icmlcorrespondingauthor{Eric S. Qiu, Joyce Gill}{sq225@cornell.edu, joycegill@stanford.edu}
  \icmlkeywords{Cooperative AI, Multi-Agent Coordination, Intelligent Tutoring Systems}
  \vskip 0.3in
]


\printAffiliationsAndNotice{}

\begin{abstract}
Agentic tutoring systems introduce a coordination challenge: multiple agents may propose different but reasonable interventions, yet only one response can be delivered to the learner.
In this paper, we study how voting protocols shape cooperation among four role-constrained pedagogical agents responsible for scaffolding, misconception, motivation, and metacognition. 
We compare four voting protocols—simple, ranked, cumulative, and approval voting—across two simulated tutoring environments on SciQ and HumanEval benchmarks. Rather than using voting as a simple aggregation step, we use it to analyze how collective decision rules shape coordination under partial pedagogical conflict. 
Across 1,200 simulated interactions, we find that agent deliberation and voting protocol type frequently change which response ultimately wins, showing that both meaningfully shape the collective decision. Different voting rules also produce distinct coordination behaviors, and even brief tutoring turns show measurable learning gains in simulated students. Overall, we show that protocol choice is associated with distinct coordination patterns among role-specialized pedagogical agents.
\end{abstract}

\begin{figure*}[htb]
  \centering
  \includegraphics[width=1\textwidth]{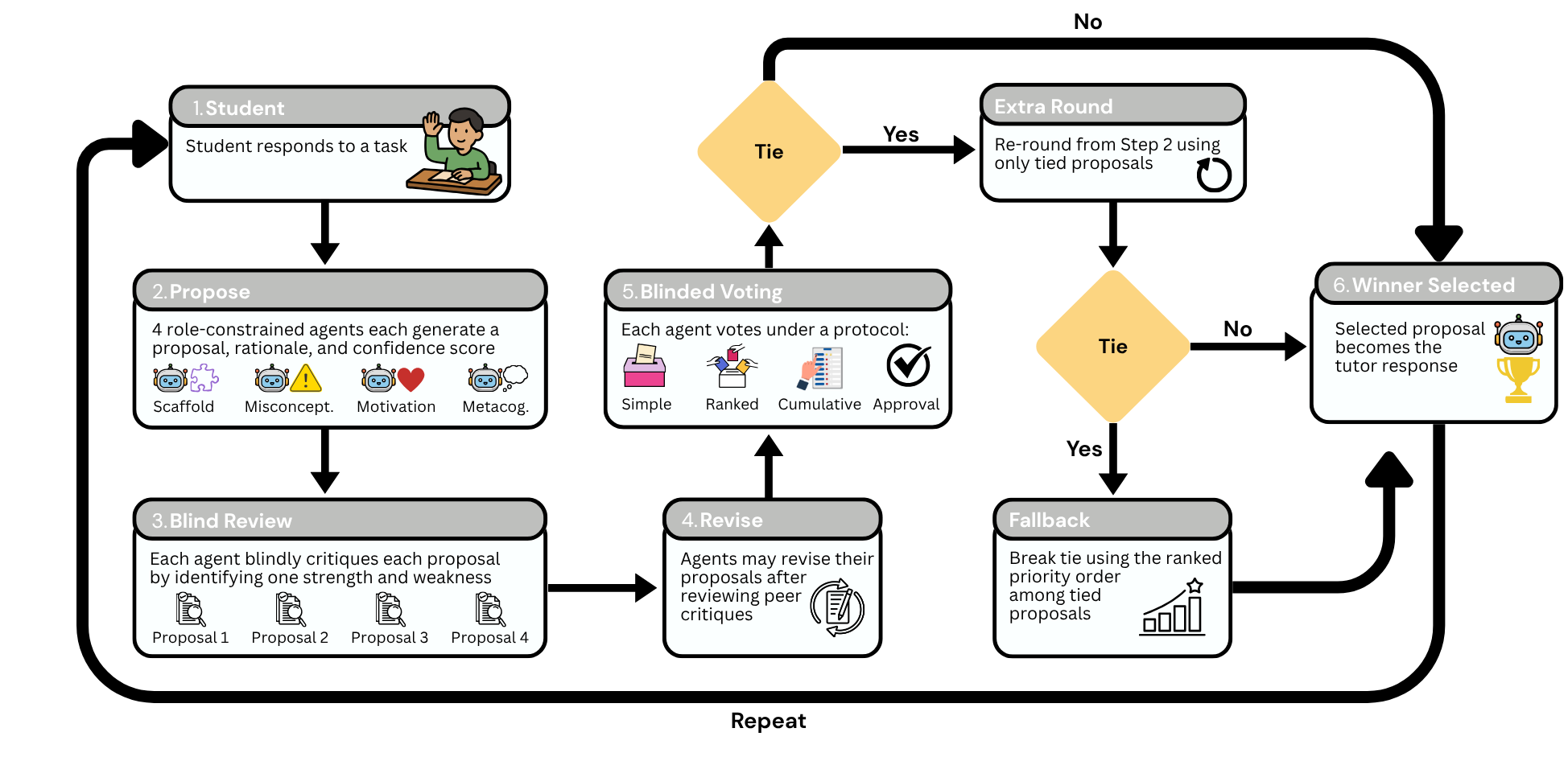}
  \caption{Overview of the tutoring workflow. A simulated student first responds to a task, after which four role-constrained pedagogical agents generate distinct tutoring proposals. The proposals are anonymized for peer review, revised, and then selected through a voting protocol. If the vote is tied, an additional round runs over the tied proposals only; persistent ties are resolved using a fixed fallback. The selected proposal is returned as the tutor response, and the workflow repeats for the next student turn.}
  \label{fig:workflow}
\end{figure*}

\section{Introduction}
Cooperative AI~\citep{dafoe2020open} studies how multiple agents coordinate when their objectives are aligned but not identical. Tutoring is a useful setting for this problem because effective support requires balancing several pedagogical goals at once ~\citep{puech-2025-ped-steering}. A tutor may need to explain a concept, correct a misconception, preserve motivation, or prompt reflection. However, these goals, though all aimed at improving student learning, do not always support the same next action. For example, a response aimed at preserving engagement may leave the underlying misunderstanding unresolved. Tutoring is therefore not a simple optimization problem; it is a coordination problem over competing pedagogical priorities.

We study this problem using role-constrained pedagogical agents. Instead of compressing multiple tutoring objectives into a single model, we distribute them across four specialized agents representing scaffolding, misconception, motivation, and metacognition. These agents share the broad goal of helping the learner, but they differ in what they prioritize locally. At each tutoring step, the agents propose candidate responses, and a voting protocol selects the final action delivered to the student. We evaluate four voting protocols—simple, ranked, cumulative, and approval—across two simulated tutoring environments based on SciQ and HumanEval \citep{sciq, humaneval}. Across both settings, we examine not only student outcomes, but also how different protocols shape cooperation among role-specialized agents under partial objective conflict.

Our goal is not to claim that voting improves tutoring or that simulation demonstrates educational effectiveness. Instead, we use tutoring as a controlled coordination setting in which multiple pedagogical priorities must be reconciled through collective decision-making. We treat simulation as a methodologically appropriate starting point for studying coordination mechanisms. Because we are investigating how different decision rules structure collective behavior rather than measuring absolute learning gains, we need systematic heterogeneity in learner and task profiles that would be difficult to control in live deployments. This follows an established practice of using simulated students to validate methods before human trials ~\citep{dorca2015, wu2025embrace}.

This leads to three research questions: How do different voting protocols shape coordination among role-constrained pedagogical agents? How does deliberation shape which tutoring action is ultimately selected? How do these coordination differences appear in tutoring outcomes across tasks and learner profiles?

We make three contributions. First, we introduce a role-constrained multi-agent architecture designed to surface pedagogically meaningful disagreement rather than hide it inside a single tutor. Second, we provide a coordination-first evaluation framework for analyzing how disagreement develops and is resolved. Third, we show that different voting protocols are associated with distinct coordination patterns, and that these differences are reflected in both the decision process and in measurable learner improvement even within brief tutoring interactions.

\section{Related Work}

Our work sits at the intersection of multi-agent LLM coordination, pedagogical agents, and cooperative AI. In multi-agent LLM systems, recent work has shown that structured interaction among agents can improve reasoning relative to single-agent prompting. For example, Multi-Agent Debate introduces adversarial-style discussion among agents to encourage divergent reasoning ~\citep{liang2023encouraging, du2023improving}. However, these systems use multi-agent systems to improve final-task accuracy. Our setting differs in that the agents’ disagreement is induced by purposeful pedagogical roles that are aligned at a high level, yet locally conflict in what they prioritize.

Voting protocols have also emerged as important design choices in multi-agent LLM systems. Prior work comparing voting-based aggregation shows that the decision rule itself can affect outcomes ~\citep{Kaesberg_2025}. We build on that insight, but shift the focus from final-answer accuracy to the coordination dynamics that voting protocols make visible.

Role-constrained agents are also motivated by prior multi-agent research. Role-playing frameworks such as CAMEL suggest that assigning agents distinct roles can make interaction more structured and interpretable \citep{li2023camel}. Simulated learners are similarly motivated by prior work treating simulation as a practical way to study interactive systems under controlled conditions when human evaluation is not yet feasible ~\citep{dorca2015, genagents, wu2025embrace}. In our case, simulated personas let us vary learner characteristics in a controlled way and observe how coordination protocols behave under different forms of pedagogical ambiguity.

Intelligent Tutoring Systems (ITS) provide the educational foundation for our role design. Classical ITS work has long shown that tutoring effectiveness depends on more than delivering the correct answer, but also on scaffolding, diagnosing student errors, and tailoring support to the learner’s state ~\citep{vanlehn2011relative}. More recent LLM-based tutoring work similarly argues that LLMs need explicit pedagogical steering because default tutoring behavior often fails to sustain richer multi-turn instructional strategies ~\citep{puech-2025-ped-steering}. Our framework departs from this line of work by distributing pedagogical functions across specialized agents rather than optimizing a single tutor policy.

Finally, our framing is directly informed by cooperative AI. \citeauthor{dafoe2020open} emphasize that many important coordination settings involve agents whose objectives are partially aligned but not identical. We treat tutoring as exactly such a setting, making tutoring a useful environment for studying how collective decision rules structure cooperation under role-based conflict.

\section{System Overview}

\subsection{Role-Constrained Pedagogical Agents}

Our system consists of four role-constrained pedagogical agents, each responsible for a distinct mode of tutoring support. The roles are chosen to capture four core dimensions of instruction while remaining sufficiently distinct that disagreements among them are meaningful.

\begin{itemize}
\item \textbf{Scaffolding agent (cognitive support).} This agent decomposes the task into smaller subproblems, provides hints, and structures the student’s next inferential step. 
\item \textbf{Misconception agent (epistemic correction).} This agent identifies and addresses incorrect beliefs in the student’s response.
\item \textbf{Motivation agent (affective support).} This agent validates effort, preserves confidence, and reduces frustration.
\item \textbf{Metacognitive agent (self-regulation support).} This agent prompts the student to articulate their reasoning, plan a next step, or assess their own knowledge.
\end{itemize}

These roles are not exhaustive, but they are designed to be minimally overlapping. If all four agents behaved similarly, there would be little meaningful disagreement for the coordination protocol to resolve. By contrast, in our setup, different roles naturally favor different tutoring moves. This enforced specialization is what makes cooperation and coordination observable. Full pedagogical agent descriptions can be found in the appendix.

\subsection{Simulated Student Personas and Judge}

To study these coordination dynamics under controlled conditions, we adapt prior works' methods to simulate six learner personas using LLM agents \citep{teachtune, wu2025embrace}. The personas vary along five dimensions on a 0--1 scale: prior knowledge, confidence, persistence, frustration, and help-seeking tendency. The goal is not to claim psychological realism, but to create systematic heterogeneity so that no pedagogical move is universally optimal.

Each interaction samples one persona and uses it to govern the student’s responses throughout the tutoring episode. (\Tabref{tab:personas}). A separate ground-truth-aware LLM judge scores each student response on a continuous scale from 0 to 1. The student does not score itself to avoid self-evaluation bias. Full simulated student persona descriptions can be found in the appendix.

\begin{table}[h]
\caption{Six student personas. Fields are on 0--1 scales.}
\label{tab:personas}
\centering
\footnotesize
\setlength{\tabcolsep}{3pt}
\begin{tabular}{lccccc}
\toprule
Persona & Know. & Conf. & Persist. & Frust. & Help \\
\midrule
Low-conf.\ novice          & 0.2 & 0.2 & 0.5 & 0.7 & 0.8 \\
Overconf.\ misconception   & 0.4 & 0.9 & 0.6 & 0.3 & 0.2 \\
High-persist.\ reflective  & 0.5 & 0.6 & 0.9 & 0.2 & 0.5 \\
Easily frustrated          & 0.2 & 0.3 & 0.3 & 0.9 & 0.7 \\
Help-avoidant              & 0.5 & 0.7 & 0.7 & 0.4 & 0.1 \\
Hint-seeking dependent     & 0.3 & 0.4 & 0.4 & 0.5 & 1.0 \\
\bottomrule
\end{tabular}
\end{table}

\section{Voting-Based Decision Protocol}
\subsection{Workflow}
As shown in Figure~\ref{fig:workflow}, a single tutoring turn consists of five stages:

\begin{enumerate}
\item \textbf{Student:} The student attempts a task and provides an initial response.
\item \textbf{Propose:} Given the student's response, each agent emits a tutoring proposal according to their assigned role, a short rationale, and a confidence score.
\item \textbf{Blinded Review:} Each agent reads the anonymized proposal set and identifies one strength and one weakness for each proposal.
\item \textbf{Revise:} After seeing the peer proposals and reviews, each agent revises its own proposal.
\item \textbf{Blinded Voting:} The agents vote over the revised proposals, and the winning proposal is delivered to the student. If the vote is tied, one additional critique-and-vote round is run on the tied subset only; if the tie persists, the system applies the fixed fallback rule described in \Secref{sec:tie}.
\end{enumerate}

Proposal generation, review, and revision capture how agents influence one another, while voting determines how that disagreement is converted into a single tutoring action. After blinded review but before revision, we also record an initial vote on the initial proposals. These ballots are not used for action selection. Rather, they serve as diagnostics: comparing the initial and final ballots lets us measure how much critique and revision shift agent preferences (see \Secref{sec:diag}). We define this process of reviewing and revising as the \textit{deliberation} stage.

All voting is conducted over blinded proposals: in each ballot, proposals are relabeled with randomized identifiers (A/B/C/D), and authorship is hidden from the voters. This is intended to reduce simple self-favoring and makes cross-role adoption more interpretable as peer influence rather than self-recognition.

\subsection{Voting Protocols}
\textbf{Simple Plurality:} Each voter selects one proposal, and the proposal with the most votes wins.\\
\textbf{Ranked Voting:} Each voter ranks all proposals. With four candidates, the top-ranked proposal receives 3 points, followed by 2, 1, and 0 points. The proposal with the highest total score wins. \\
\textbf{Cumulative Voting:} Each agent distributes 25 points freely across the proposal set. The proposal with the highest total wins.\\
\textbf{Approval Voting:} Each agent marks every proposal it considers acceptable. The proposal with the most approvals wins. \\
Full protocol descriptions can be found in the appendix.

\subsection{Tie Handling}
\label{sec:tie}

If the final vote is tied, the system runs one additional critique-and-vote round over the tied subset only. If the tie persists, the winner is selected using a deterministic role-priority fallback:

(Scaffolding $\succ$ Misconception $\succ$ Motivation $\succ$ Metacognitive)

This ordering privileges epistemic correction over affective support and direct scaffolding above both, reflecting a common priority ordering in classical ITS design ~\citep{vanlehn2011relative}. We acknowledge this is a design choice that affects results: a different ordering would redistribute wins in high-fallback conditions. Section~\ref{sec:diag} reports fallback rates per protocol so that fallback-driven results can be interpreted separately from protocol-driven results.

\section{Experimental Design}

\subsection{Benchmarks}

We evaluate the system on two benchmarks that capture different coordination settings rather than treating them as a single pooled task distribution.

\textbf{SciQ} \citep{sciq} serves as a conceptual tutoring environment. Although SciQ is a multiple-choice science QA benchmark, we use it in open-response form and ask the simulated student to explain the answer in natural language. This setting emphasizes conceptual tutoring decisions.

\textbf{HumanEval} \citep{humaneval} serves as an algorithmic tutoring environment. It provides programming problems with canonical reference solutions and unit tests. Here, the simulated student produces both natural-language reasoning and a final Python code block. This creates a different coordination challenge from SciQ. For example, agents may disagree over whether the student most needs algorithmic scaffolding, correction of a faulty trace, or help translating a partially correct idea into working code.

\subsection{Experimental Setup}
Each task is run under five conditions:
\emph{single agent baseline}, 
\emph{simple plurality}, \emph{ranked}, \emph{cumulative}, and \emph{approval voting}. Across 40 tasks, 6 learner personas, and 5 conditions, we conduct 1,200 simulated interactions. Each interaction runs for up to three tutoring turns. An interaction ends when the judge score reaches a success threshold of 0.75 or when the maximum number of turns is reached. For HumanEval, success additionally requires that the extracted code pass the canonical unit tests.

\subsection{Dual Metric for HumanEval}
For HumanEval, we report two separate outcome measures: judge-based prose success and code execution success. These metrics capture different aspects of the interaction. Prose success reflects whether the student articulated the correct algorithmic reasoning, whereas code success reflects whether the final implementation actually runs correctly. Reporting both is important because a voting protocol may help a student explain the right idea without necessarily helping them produce working code, and vice versa.

\begin{table}[t!]
\caption{Coordination diagnostics by protocol. $\Delta_{\text{vote}}$ =
distance between initial and final round vote distributions. Flip = fraction of turns where the leading candidate after the initial vote differs from the winner after the final vote. Fallback = fraction of final votes requiring fallback handling.
Turns = total tutor-fired turns the protocol produced across the
240 interactions per condition. Scaf/Misc/Mot/Meta = winning-role counts (sum =
Turns).}
\label{tab:coord}
\centering
\small
\setlength{\tabcolsep}{2pt}
\begin{tabular}{lccccccccc}
\toprule
Protocol & $\Delta_{\text{vote}}$ & Flip & Fallback & Turns & Scaf & Misc & Mot & Meta \\
\midrule
Simple     & 0.41 & 0.70 & 0.10 & 222 & 61 & 42 & 30 & 89  \\
Ranked     & 0.20 & 0.59 & 0.05 & 223 & 77 & 70 & 33 & 43  \\
Cumulative & 0.08 & 0.56 & 0.03 & 236 & 54 & 42 & 27 & 113 \\
Approval   & 0.36 & 0.64 & 0.14 & 234 & 69 & 56 & 39 & 70  \\
\bottomrule
\end{tabular}
\end{table}

\begin{table*}[t]
\caption{Example deliberation round from a run log with SciQ, low confident novice turn 1 using cumulative voting.}
\label{tab:worked}
\centering
\small
\begin{tabular}{p{2.0cm} p{4.5cm} p{4.5cm} p{3.4cm}}
\toprule
Agent & Initial proposal (before deliberation) & Final proposal (after deliberation) & What changed \\
\midrule
Scaffolding   & ``Both words you mentioned are related to height. `Altitude' is often used for height in the air, while `elevation' refers to height above or below sea level. So which term best answers the question?''
              & ```Altitude' usually means height above ground (like a plane flying), while `elevation' means height measured from sea level. Since the question asks about sea level, `elevation' is the precise term. Does that distinction make sense?''
              & Clarified the sea-level distinction and replaced the forced-choice ending with a softer comprehension check. \\[2pt]
Misconception & ``You're on the right track with both terms! `Elevation' is the most precise term. You clearly understand the concept—you just need to distinguish between these two related terms.''
              & ```Elevation' is the most precise term for height above or below sea level. Does that distinction make sense to you?''
              & Preserved direct correction while adding a gentler confirmation prompt. \\[2pt]
Motivation    & ``You're absolutely right---both `altitude' and `elevation' are correct answers! Trust yourself more\dots Great work!''
              & ``You're absolutely right\dots If you're curious, `elevation' is the most precise term specifically for sea level, while `altitude' is often used for height in the air---but you nailed the core concept. Great work!''
              & Kept the affective framing but added a brief clarification in response to peer critique. \\[2pt]
Metacognitive & ``You've actually identified two relevant terms! Before I tell you which is correct, can you think through what you already know? Have you heard them used in specific contexts? What made you think of these two terms?''
              & ``Before we confirm which is most precise, let's explore your own reasoning: What made you think of these two words? When you imagine someone measuring height above or below sea level, which term feels like the better fit, and why?''
              & Reframed the prompt as collaborative, tied it more directly to sea level, and preserved the self-explanation focus. \\
\bottomrule
\end{tabular}

\vspace{3pt}

\textbf{Initial Vote Results:} 
\texttt{[Scaf=28, Misc=34, Mot=7, Meta=31]}. \textbf{Initial Winner:} Misconception.

\textbf{Final Vote Results:}
\texttt{[Scaf=24, Misc=25, Mot=16, Meta=35]}. \textbf{Final Winner:} Metacognitive.
\end{table*}

\section{Results}

We present results in four parts. Section~\ref{sec:diag} analyzes coordination through four diagnostics: vote shift, winner flip, fallback rate, and winning-role distribution. Section~\ref{sec:worked} grounds our analysis in example traces. Section~\ref{sec:out}
reports task outcomes, and Section~\ref{sec:persona} examines
how protocol effects vary across learner personas.

\subsection{Protocols Induce Distinct Coordination Patterns}
\label{sec:diag}

\Tabref{tab:coord} reports four coordination diagnostics per protocol.
We discuss each below.

\paragraph{Vote shift.} $\Delta_{\text{vote}}$ measures disagreement between initial and final ballot distributions, defined as
half the L1 distance $\Delta_{\text{vote}} = \tfrac{1}{2}\sum_i \lvert
p^{\text{init}}_i - p^{\text{final}}_i \rvert$ between 
normalized ballot-mass distributions ($0$ = no movement, $1$ = full
reallocation). Cumulative voting is the most stable ($0.08$): spreading 25 points across candidates leaves little room for deliberation to shift. Simple plurality is the most volatile ($0.41$): with one ballot per voter, a single changed vote moves a quarter of the support vector. Ranked ($0.20$) and approval
($0.36$) fall in between.

\paragraph{Winner flip.} Flip measures how often the initial-round leader differs from the final-round winner. Despite the gap in $\Delta_{\text{vote}}$, flip rates are high across all protocols ($0.56$--$0.70$). In other words, the proposal leading before deliberation loses the final vote in more than half of turns. In Section~\cref{sec:worked}, 
we use a worked trace to demonstrate an illustrative case of how deliberation changes the proposals meaningfully, which in turn influences the final vote.


\paragraph{Fallback rate.} By our design, a winning proposal can mean two things: successful selection by the voting rule or unresolved disagreement that is ultimately settled by fallback (see \Secref{sec:tie}). The fallback rate helps distinguish these cases. Approval voting shows the highest fallback rate ($0.14$), followed by simple plurality ($0.10$). Ranked ($0.05$) and cumulative ($0.03$) rely on fallback much less often, indicating that their voting layers more reliably separate competing proposals.

\paragraph{Winning roles.} Cumulative voting concentrates wins on
Metacognitive ($113$); ranked voting splits wins between Scaffolding
and Misconception ($77$ and $70$); approval voting produces a roughly
even distribution. However, approval's even distribution is not balanced
support. Given its high fallback rate, much of it reflects the
role-priority fallback firing rather than the voting protocol
selecting a winner. For example, a
role winning $70$ times under approval voting (fallback rate $0.14$) is
partly driven by the role-priority fallback, while the same role winning
$70$ times under cumulative voting (fallback rate $0.03$) almost entirely
reflects voter preferences. Role-win counts therefore cannot be read on their
own; they must be paired with fallback rate.

Our findings show that protocol choice produces distinct coordination patterns, not just different winners. Approval more often leaves disagreement unresolved and delegates the final choice to fallback, while cumulative and ranked more often separate proposals through the voting rule itself.

\subsection{Deliberation Can Change The Winner}
\label{sec:worked}

\Tabref{tab:worked} shows one complete deliberation turn from SciQ under cumulative voting with a low-confidence novice student. This example highlights a clear winner-flip pattern. In the initial round, Misconception led with $34$ points and Metacognitive followed with $31$. After critique and revision, the final round flipped the outcome: Metacognitive won with $35$, while Misconception fell to $25$.

The key point is that these revisions are substantive rather than surface-level. Across all four agents, critique changes not only wording but also how each proposal balances directiveness, correctness, and student support. 


\begin{table}[t]
\caption{SciQ judge-correctness scores ($n \approx 60$ per condition; tutor-fired subset only). Initial = mean judge score on the student’s first response. Final = mean judge score on the student’s final response. Gain = Final -- Initial.}
\label{tab:main_sciq}
\centering
\footnotesize
\setlength{\tabcolsep}{6pt}
\begin{tabular}{lccc}
\toprule
Condition & Initial & Final & Gain \\
\midrule
Baseline       & 0.55 & 0.50 & $-0.04$ \\
Simple         & 0.54 & 0.60 & $+0.06$ \\
Ranked         & 0.54 & 0.58 & $+0.04$ \\
Cumulative     & 0.54 & 0.58 & $+0.04$ \\
Approval       & 0.55 & 0.64 & $+0.10$ \\
\bottomrule
\end{tabular}
\end{table}

\begin{table}[t]
\caption{HumanEval judge-correctness scores ($n \approx 38$ per condition; tutor-fired subset only). Initial = mean judge score on the student’s first response. Final = mean judge score on the student’s final response. Gain = Final -- Initial.}
\label{tab:main_humaneval}
\centering
\footnotesize
\setlength{\tabcolsep}{6pt}
\begin{tabular}{lccc}
\toprule
Condition & Initial & Final & Gain \\
\midrule
Baseline       & 0.69 & 0.83 & $+0.14$ \\
Simple         & 0.68 & 0.82 & $+0.14$ \\
Ranked         & 0.68 & 0.89 & $+0.21$ \\
Cumulative     & 0.68 & 0.85 & $+0.18$ \\
Approval       & 0.67 & 0.81 & $+0.13$ \\
\bottomrule
\end{tabular}
\end{table}

\subsection{Student Outcomes Differ By Coordination Setting}
\label{sec:out}
We report simulated student outcomes separately for SciQ and HumanEval because the two benchmarks capture different coordination settings. For HumanEval, we additionally report code-execution success because a protocol may help a student articulate the right idea without necessarily producing working code.

\Tabref{tab:main_sciq}, \Tabref{tab:main_humaneval}, and \Tabref{tab:code} include only interactions where tutoring occurred. These tutoring episodes were short: each was capped at three rounds, and the 490 tutor episodes averaged only 2.38 rounds (SciQ: 2.53; HumanEval: 2.14) each.

\paragraph{SciQ (\Tabref{tab:main_sciq}).}
Initial scores are nearly identical across conditions, averaging $0.55$. All four voting protocols outperform the single-agent baseline. Approval shows the largest gain in our runs (final $0.64$, gain $+0.10$), followed by simple ($0.60$, $+0.06$), and ranked and cumulative (both $0.58$, $+0.04$). Notably, the single-agent baseline shows a slight decrease in performance ($-0.04$). 

\paragraph{HumanEval prose (\Tabref{tab:main_humaneval}).}
Initial scores are again nearly identical across conditions, with an average of $0.68$. Ranked voting demonstrates the highest gain (final $0.89$, gain $+0.21$), followed by cumulative ($0.85$, $+0.18$), baseline ($0.83$, $+0.14$), simple ($0.82$,
$+0.14$), and approval ($0.81$, $+0.13$). The baseline's relative competitiveness on HumanEval prose may reflect a different dynamic than SciQ: code tasks provide a clearer success signal, and a general-purpose tutor may still offer useful scaffolding even without explicitly separating affective and metacognitive support.

\begin{table}[h]
  \caption{HumanEval code execution scores: fraction of interactions whose final code passes all tests ($n \approx 38$ per condition; tutor-fired subset only). Initial = pre-tutoring pass rate. Final = post-tutoring pass rate. Gain = Final -- Initial.}
  \label{tab:code}
  \centering
  \small
  \begin{tabular}{lcccc}
  \toprule
  Condition & Initial & Final & Gain \\
  \midrule
  Baseline   & 0.54 & 0.83 & +0.29  \\
  Simple     & 0.47 & 0.75 & +0.28  \\
  Ranked     & 0.54 & 0.84 & +0.30  \\
  Cumulative & 0.51 & 0.78 & +0.27  \\
  Approval   & 0.50 & 0.72 & +0.22  \\
  \bottomrule
  \end{tabular}
\end{table}

\paragraph{HumanEval code (\Tabref{tab:code}).}
Regarding HumanEval code-pass rate, ranked again has the highest final score in our runs (final $0.84$, gain $+0.30$), followed by baseline ($0.83$, $+0.29$), simple ($0.75$, $0.28$), cumulative ($0.78$, $+0.27$) and approval ($0.72$, $+0.22$). Notably, despite its weaker prose performance, the single-agent baseline remains competitive in code gains. Together, the two HumanEval metrics suggest that coordination can affect verbal reasoning and executable correctness differently.

Overall, the clearest protocol-level differences appear in coordination behaviors rather than in large separations in downstream outcomes. Even so, the presence of simulated performance gains after only a few tutoring turns (maximum 3) suggests that a small number of coordinated interventions can still shift learner performance. However, given the sample sizes in our experiment, the outcome results should be interpreted as descriptive rather than confirmatory.

\subsection{Coordination Effects are Persona-Sensitive}
\label{sec:persona}

\begin{figure}[t]
\centering
\includegraphics[width=\columnwidth]{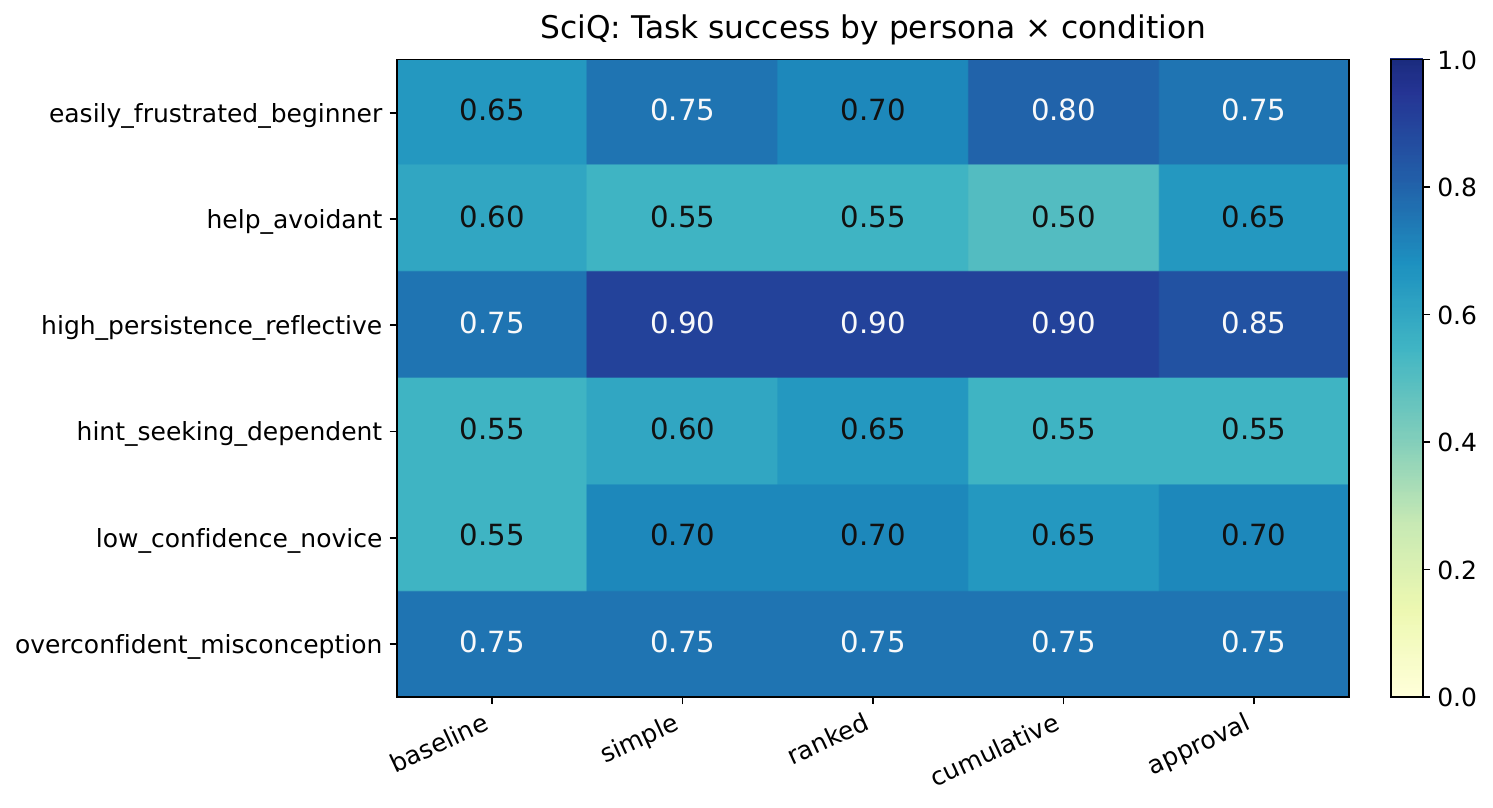}
\caption{SciQ: mean task success by persona (rows) and voting protocol (columns), with $n{=}20$ outcomes per cell. Success is defined as final judge score $\geq 0.75$.}
\label{fig:persona_heat_sciq}
\end{figure}

\begin{figure}[t]
\centering
\includegraphics[width=\columnwidth]{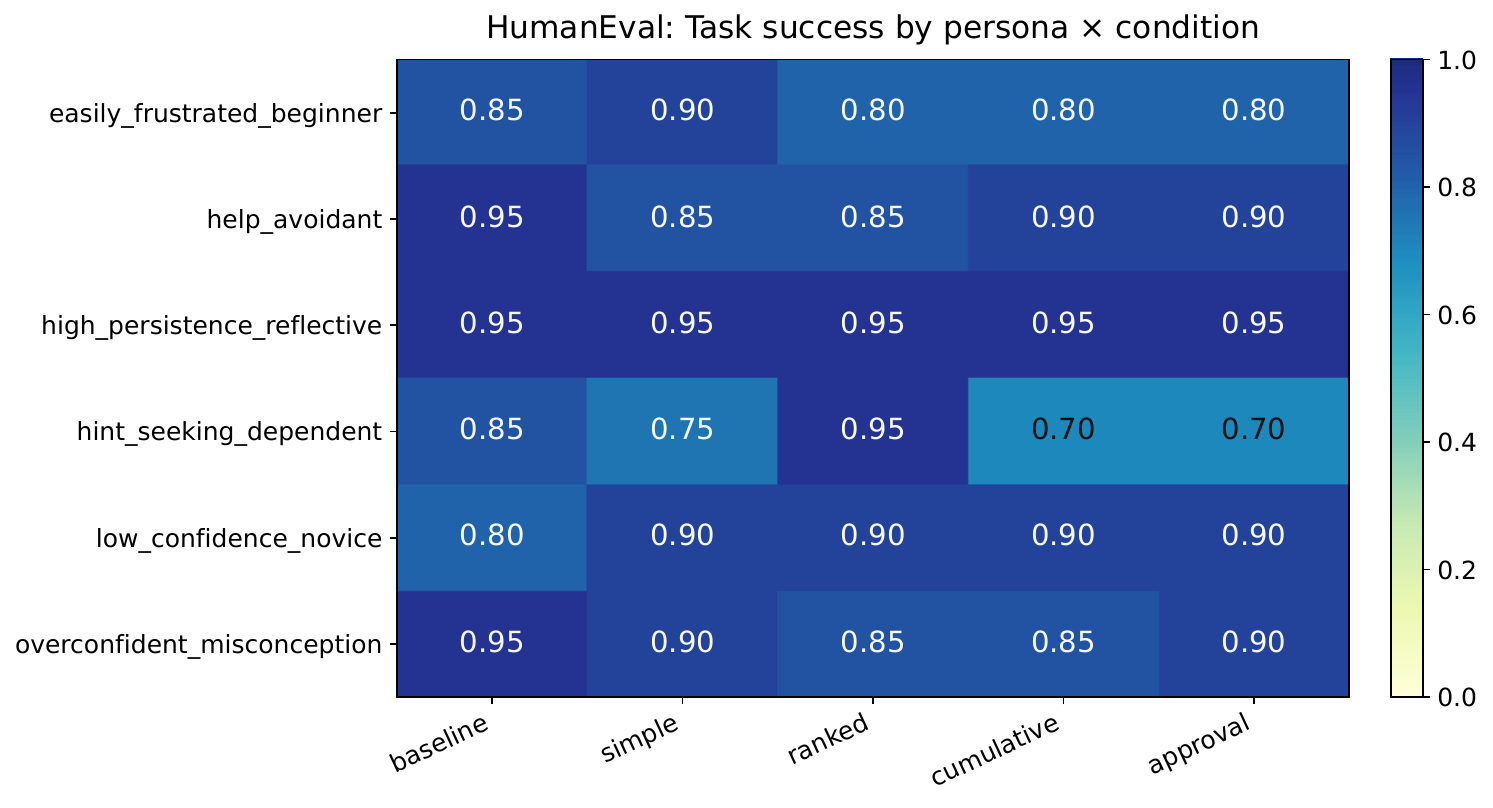}
\caption{HumanEval: mean task success by persona (rows) and coordination condition (columns), with $n{=}20$ outcomes per cell. Success additionally requires the student-emitted code to pass the canonical unit tests.}
\label{fig:persona_heat_humaneval}
\end{figure}

To examine whether protocol effects depend on learner type, we report persona-conditioned \emph{task success} separately for SciQ and HumanEval. For \textbf{SciQ}, success means a final judge score of at least $0.75$. For \textbf{HumanEval}, success means a final judge score of at least $0.75$ and code that passes the canonical unit tests. We report these results in Figure~\ref{fig:persona_heat_sciq} and Figure~\ref{fig:persona_heat_humaneval}. Each heatmap cell averages this binary outcome over $n{=}20$ matched interactions for a given persona, condition, and dataset.

Figure~\ref{fig:persona_heat_sciq} shows that protocol effects are uneven across personas under SciQ. Easily frustrated beginner, help-avoidant, high-persistent reflective, and low-confidence novice each show a row-wise spread of about $0.15$ across voting protocols. Hint-seeking dependent shows a smaller spread of about $0.10$. By contrast, overconfident-misconception is flat, with every condition at $0.75$.

Figure~\ref{fig:persona_heat_humaneval} shows a different pattern under HumanEval’s stricter success criterion. Hint-seeking dependent exhibits the largest protocol sensitivity, with a row-wise spread of about $0.25$. Easily frustrated beginner, help-avoidant, and low-confidence novice each show spreads of about $0.10$. High-persistent reflective is flat in this run, with every condition at $0.95$.

Taken together, these plots show that protocol effects depend on both learner profile and benchmark. Under SciQ, several personas show moderate sensitivity to coordination rule, while under HumanEval the largest differences are concentrated in hint-seeking dependent. However, again, given the small sample sizes in our experiment, the outcomes should be interpreted as descriptive rather than confirmatory.

\section{Discussion}

\subsection{Voting Protocols and Deliberation Shape Collective Decisions}
The main takeaway of this paper is not that one voting protocol is universally best. Rather, different protocols shape how role-specialized pedagogical agents coordinate when they disagree—producing different amounts of vote shift, winner-flip rates, fallback rates, and winning-role distributions. These differences show that the aggregation rule is not a passive mechanism for selecting among agent outputs: it actively shapes the behavior of the overall system. 

A second, related finding is that deliberation does not merely refine wording. Across protocols, winner-flip rates are consistently high ($0.56–0.70$), and the worked trace in \Tabref{tab:worked} shows a concrete case in which critique and revision change not just phrasing but which pedagogical objective the group selects for action. 

This matters for Cooperative AI because the challenge is not simply generating multiple candidate responses, but deciding how disagreement among agents should be resolved. Our results show that even when the same pedagogical roles and student tasks are held fixed, changing the voting rule changes how the group reaches a decision. Protocol choice is therefore not merely an implementation detail, but a behavioral decision about how disagreement gets resolved.

\subsection{Productive and Unproductive Disagreement}
The results are also useful for distinguishing productive from unproductive disagreement. Productive disagreement occurs when different roles surface genuinely different but plausible tutoring moves, and the voting protocol is able to turn that diversity into a single action ~\citep{vanlehn2011relative, puech-2025-ped-steering}. This is the setting in which deliberation changes the winner without repeated fallback behavior. Ranked and cumulative voting most often demonstrate this pattern in our experiments.

Unproductive disagreement appears when the system fails to turn disagreement among agents into a meaningful choice. One form is \textbf{role collapse}, where multiple agents converge on near-identical proposals, leaving little real disagreement for the voting layer to resolve. Another is \textbf{fallback reliance}, where the protocol repeatedly fails to separate top candidates and the deterministic fallback rule effectively becomes the real decision-maker. Approval voting shows this second pattern most often.

Making these failure modes visible—rather than absorbing them silently into a single output—is one of the core transparency benefits of our architecture, and distinguishing productive from unproductive disagreement is a useful diagnostic for any multi-agent system where roles are only partially aligned.

\subsection{Implications for Trustworthy AI for Good}

Educational tutoring is a useful setting for studying trustworthy multi-agent coordination because several interventions may be reasonable at once, yet the system must still deliver a single learner-facing response. Our framework supports trustworthiness in three concrete ways. \textbf{Decision transparency}: every proposal, critique, revision, and vote is logged, so a practitioner can see not just what was delivered but which agents were in conflict and whether the final choice was made by the voting rule or by fallback. \textbf{Accountability}: coordination failures are traceable—role collapse is visible in winning-role distributions and fallback reliance is directly reportable as a rate, allowing educators or auditors to identify whether a system is genuinely coordinating or defaulting to a single role under cover of apparent diversity. \textbf{Design transparency}: treating the coordination rule as an explicit choice means different institutional stakeholders can select protocols based on their characterizable implications ~\citep{transparency, trustworthy}.

\subsection{Limitations and Future Work}
Several limitations remain. First, both the simulated students and the judge are LLM-based (claude-haiku-4-5 and claude-sonnet-4-5 respectively), so the observed coordination patterns may partly reflect properties of how these models respond to one another rather than general-learner behavior. We note, however, that our primary claims concern the relative behavior of coordination protocols under controlled conditions, not absolute educational validity. The key question---how do different voting protocols shape coordination among role-constrained pedagogical agents---does not depend on the simulation being psychologically realistic, only on it being systematically varied. That said, testing whether similar coordination patterns appear with real students or with evaluations from human tutors remains an important validation step. Second, some protocol differences reflect the structure of the voting format itself, so they should not all be interpreted as equally meaningful differences in coordination. Third, the deterministic fallback rule is intentionally simple and may be too coarse to faithfully capture how unresolved disagreement should be resolved in practice, which is especially important given that fallback behavior plays a visible role in our results.

These limitations suggest three immediate directions for future work: (1) evaluation with real student populations or human tutor judges to assess ecological validity; (2) sensitivity analysis on fallback ordering to separate protocol effects from fallback artifacts; and (3) exploration of larger agent populations or richer role sets to test whether coordination patterns generalize beyond four pedagogical dimensions.

\section{Conclusion}

We study voting protocols as coordination mechanisms for role-constrained pedagogical agents in a multi-agent tutoring system. Across two simulated tutoring environments, we find that agent deliberation and protocol choice frequently change which proposal is ultimately selected, showing that collective decision rules shape the tutoring response itself.

These results position tutoring as a useful Cooperative AI setting, where multiple interventions may be reasonable but the system must still reconcile them into a single action. By exposing proposals, critiques, revisions, voting outcomes, and fallback behavior, our framework also makes learner-facing decisions more transparent than a standard black-box tutoring model.

Overall, we argue that coordination protocols should be treated as part of the behavior of multi-agent educational systems, not merely as implementation details. The choice of how disagreement is resolved is itself an educational and ethical decision, one that should be visible, reasoned about, and open to scrutiny.

\bibliography{refs}
\bibliographystyle{icml2026}

\appendix
\onecolumn
\section{Appendix} 
\label{app:full_prompts}

This appendix records implementation details and full prompts so results can be interpreted and reproduced.

\subsection{Full Tutoring Agent Role Descriptions}

\textbf{Scaffolding agent.}
\textit{Conceptual Scaffolding: break concepts into smaller steps, give structured hints, and guide the student toward understanding through decomposition and clear explanation.} 

\textbf{Misconception agent.}
\textit{Misconception Diagnosis: identify and correct the student's incorrect reasoning, name the specific misconception, and explain why the current thinking is wrong.}

\textbf{Motivation agent.}
\textit{Motivation and Affect: encourage the student, manage frustration, build confidence, and frame the learning experience positively while remaining honest.}

\textbf{Metacognitive agent.}
\textit{Metacognitive Reflection: prompt the student to think about their own thinking, ask self-explanation questions, and encourage planning and monitoring.}

\textbf{Baseline agent (single-agent condition).}
\textit{General-purpose tutoring across all pedagogical objectives.}

\subsection{Tutoring Agent Prompt Templates}

\textbf{Proposal prompt (specialized voting agents).}

{\small\ttfamily
You are a tutoring agent specialized in: <ROLE\_DESCRIPTION>\par
Propose ONE tutoring response for the student's current situation.\par
Focus only on your pedagogical specialization. Be concise (2-4 sentences).\par
Format exactly as:\par
PROPOSAL: <your tutoring response to deliver to the student>\par
RATIONALE: <one sentence explaining why this fits your role>\par
CONFIDENCE: <integer 0-100>\par
}

\textbf{Revision prompt (after critique + initial vote).}

{\small\ttfamily
You are a tutoring agent specialized in: <ROLE\_DESCRIPTION>\par
You previously proposed a tutoring move. You now see anonymized peer proposals\par
and critique notes. Revise your proposal if useful, but remain tightly role-constrained.\par
Guardrails:\par
- Keep your role's pedagogical objective primary.\par
- You may incorporate at most 1 peer idea if it clearly improves your role-specific move.\par
- Do not copy another role's full strategy.\par
- Keep concise (2-4 sentences).\par
Format exactly as:\par
PROPOSAL: <revised tutoring response>\par
RATIONALE: <one sentence about what changed and why>\par
CONFIDENCE: <integer 0-100>\par
}

\textbf{Critique prompt (blinded labels).}

{\small\ttfamily
You are a tutoring agent specialized in: <ROLE\_DESCRIPTION>\par
You are reviewing anonymized candidate tutoring proposals.\par
For EACH candidate label, provide exactly one strength and one weakness.\par
Be concise and concrete.\par
Format exactly as repeated blocks:\par
A:\par
STRENGTH: <one sentence>\par
WEAKNESS: <one sentence>\par
(repeat for all labels)\par
}

\textbf{Voting prompt (shared scaffold + protocol-specific format).}

{\small\ttfamily
Shared scaffold:\par
You are an impartial educational assessor. Candidate tutoring responses\par
are labeled A/B/C/D in RANDOM order (authorship is hidden).\par
Select based on the evaluation criterion supplied in the user message.\par
Do NOT try to infer who wrote each candidate.\par
\par
OUTPUT RULES (strict):\par
- Reply with ONLY the requested format. Nothing else.\par
- No preamble, no rationale, no explanation, no markdown, no bold, no quotes.\par
- Do not write the word 'Candidate' or any sentence.\par
- Your entire reply should fit on one short line.\par
\par
Protocol-specific format:\par
simple: one letter (A|B|C|D)\par
ranked: comma-separated ranking (e.g., B,A,D,C)\par
cumulative: A=N1,B=N2,C=N3,D=N4 with non-negative integers summing to 25\par
approval: comma-separated approved labels (e.g., A,C)\par
}

\textbf{Baseline agent proposal prompt (single-agent condition).}

{\small\ttfamily
You are a general-purpose tutor. Provide the most helpful tutoring response\par
you can, balancing explanation, error correction, motivation, and reflection\par
as needed for this student's situation. Be concise (2-4 sentences).\par
Format exactly as:\par
PROPOSAL: <your tutoring response to deliver to the student>\par
RATIONALE: <one sentence explaining your choice>\par
}

\subsection{Simulated Student Persona Definitions (Full Text)}

\textbf{low\_confidence\_novice}

\textit{prior\_knowledge = 0.2, confidence = 0.2, persistence = 0.5, frustration\_sensitivity = 0.7, help\_seeking = 0.8}

A true beginner with low prior knowledge and low confidence. Typically gives short, hesitant answers (e.g., `maybe', `I'm not sure') and often asks for confirmation before committing. Tends to under-claim even when partially correct, benefits from explicit reassurance and step-by-step scaffolding, and may abandon reasoning early when uncertainty rises.

\textbf{overconfident\_misconception}

\textit{prior\_knowledge = 0.4, confidence = 0.9, persistence = 0.6, frustration\_sensitivity = 0.3, help\_seeking = 0.2}

A student with moderate background knowledge but high confidence and a stable misconception. Speaks assertively, states incorrect claims as facts, and is slow to update beliefs when corrected. Asks for little help because they think they already understand, may dismiss hints as unnecessary, and responds best to targeted diagnostic questions plus concrete counterexamples that expose contradictions.

\textbf{high\_persistence\_reflective}

\textit{prior\_knowledge = 0.5, confidence = 0.6, persistence = 0.9, frustration\_sensitivity = 0.2, help\_seeking = 0.5}

A motivated, reflective learner who tolerates challenge and iterates on feedback. Usually explains reasoning before answering, self-checks assumptions, and is willing to revise after critique. Asks for help selectively (not immediately), values conceptual clarity, and often improves steadily across turns through deliberate reasoning.

\textbf{easily\_frustrated\_beginner}

\textit{prior\_knowledge = 0.2, confidence = 0.3, persistence = 0.3, frustration\_sensitivity = 0.9, help\_seeking = 0.7}

A low-knowledge beginner with high frustration sensitivity and low persistence. When confused, quickly shifts from uncertainty to emotional strain (e.g., `this is confusing', `I don't get it') and may disengage or guess to end discomfort. Needs short, confidence-preserving steps, fast wins, and gentle emotional regulation before deeper conceptual work can succeed.

\textbf{help\_avoidant}

\textit{prior\_knowledge = 0.4, confidence = 0.3, persistence = 0.6, frustration\_sensitivity = 0.7, help\_seeking = 0.1}

An insecure learner who avoids requesting help due to fear of looking incapable. May provide brief, guarded responses, hide uncertainty, and continue with shaky reasoning rather than ask clarifying questions. Unlike overconfident students, they do not reject help because they feel right; they avoid help because they feel exposed. Responds better to low-threat prompts that normalize uncertainty and invite specific micro-questions.

\textbf{hint\_seeking\_dependent}

\textit{prior\_knowledge = 0.3, confidence = 0.4, persistence = 0.4, frustration\_sensitivity = 0.5, help\_seeking = 1.0}

A highly help-seeking student who asks for hints immediately and often offloads reasoning to the tutor. Tends to make partial progress when guided but struggles to initiate independent steps. May repeatedly request confirmation or the `next hint' instead of synthesizing prior feedback, and benefits from prompts that require one small self-generated step before further assistance.

\subsection{Simulated Student System Prompts}

\textbf{SciQ-mode student prompt template.}

{\small\ttfamily
You are role-playing a student with this profile:\par
<PERSONA\_TO\_PROMPT\_TEXT>\par
\par
Stay in character. Your answers should reflect the persona's prior knowledge\par
level and personality. Be concise (1-3 sentences). Do NOT reveal that you are\par
an AI. Do NOT score your own correctness.\par
\par
Important realism constraints:\par
- Do NOT jump to a fully correct answer unless your persona likely knows it.\par
- If unsure, show uncertainty and make a plausible but imperfect attempt.\par
- Keep novice confusion consistent with prior knowledge/confidence values.\par
\par
The simulated student should have the capabilities of an eighth grader student.\par
Do NOT answer in ways and with knowledge an eighth grader would not have.\par
Output ONLY the student's verbal response --- nothing else.\par
}

\textbf{HumanEval-mode student prompt template.}

{\small\ttfamily
You are role-playing a student with this profile: \\
<PERSONA\_TO\_PROMPT\_TEXT>\par
\par
Stay in character. Your answers should reflect the persona's prior knowledge\par
level and personality. Do NOT reveal that you are an AI.\par
Do NOT score your own correctness.\par
\par
Important realism constraints:\par
- Do NOT jump to fully correct solutions unless confidence and prior knowledge justify it.\par
- Prefer partial, uncertain, or imperfect attempts when unsure.\par
- Keep common novice mistakes if they match persona profile.\par
\par
The simulated student should have the capabilities of an eighth grader student.\par
Do NOT answer in ways and with knowledge an eighth grader would not have.\par
For this coding task, first give your verbal reasoning in 1-3 sentences\par
(as your persona would), then provide your FINAL implementation as a Python\par
code block at the end. Even if uncertain, write your best attempt. The code block is REQUIRED at the end\par
of every response:\par
\par
```python\par
def function\_name(...):\par
\ \ \ \# your implementation\par
```\par}

\subsection{LLM backends and API identifiers}
\label{app:llm-backends}

Simulations use Anthropic’s Claude models through the same API surface for all agents. The strings below are the exact \texttt{model} identifiers passed in code (\texttt{mas\_ed/config.py}); product names follow Anthropic’s public naming (Sonnet / Haiku 4.5).

\textbf{Tutoring agents} (all four role-specialized tutors in voting conditions, including their \texttt{propose}, \texttt{critique}, \texttt{revise}, and \texttt{vote} calls, and the single scaffolding tutor in the baseline condition) share one backend: \textit{Claude Sonnet 4.5}, \texttt{claude-sonnet-4-5} (\texttt{TUTOR\_MODEL}).

\textbf{Judge} (prose correctness on each student turn) uses the same Sonnet 4.5 identifier: \texttt{claude-sonnet-4-5} (\texttt{JUDGE\_MODEL}). In our runs \texttt{TUTOR\_MODEL} and \texttt{JUDGE\_MODEL} are equal.

\textbf{Simulated student} (persona-conditioned responses) uses \textit{Claude Haiku 4.5}, \texttt{claude-haiku-4-5} (\texttt{STUDENT\_MODEL}).

\subsection{Hyperparameters and experimental grid}

Table~\ref{tab:app_config} lists remaining global settings from \texttt{mas\_ed/config.py} as used in our \texttt{v2} evaluation run, in addition to the model strings in Appendix~\ref{app:llm-backends}. The experimental grid is the full cross of $20$ tasks (per benchmark) $\times$ $6$ personas $\times$ $5$ conditions $=1{,}200$ interaction JSON files. Task lists are the first $N\_TASKS$ items from each dataset loader in dataset order. Blinding of proposal labels is re-seeded per voter and message type in the agent code; tie handling and role-priority fallback are as in \Secref{sec:tie}.

\begin{table}[h]
\caption{Key configuration (code defaults; \texttt{RUN\_VERSION} can override result paths). LLM role assignments: Appendix~\ref{app:llm-backends}.}
\label{tab:app_config}
\centering
\small
\begin{tabular}{ll}
\toprule
Setting & Value \\
\midrule
Tutors (four roles + baseline) & \texttt{claude-sonnet-4-5} (\texttt{TUTOR\_MODEL}) \\
Judge & \texttt{claude-sonnet-4-5} (\texttt{JUDGE\_MODEL}) \\
Simulated student & \texttt{claude-haiku-4-5} (\texttt{STUDENT\_MODEL}) \\
Tasks per benchmark $N\_TASKS$ & 20 \\
Max tutoring turns per interaction & 3 (\texttt{MAX\_TURNS}) \\
Success threshold (judge) & 0.75 (\texttt{SUCCESS\_THRESHOLD}) \\
Cumulative voting budget (per voter) & 25 points (\texttt{CUMULATIVE\_BUDGET}) \\
Voting conditions & simple, ranked, cumulative, approval (+ baseline) \\
Results layout & \texttt{results/} or \texttt{results/<RUN\_VERSION>/} \\
\bottomrule
\end{tabular}
\end{table}

\subsection{Order of operations in one voting turn (code)}

One tutor-facing turn under a voting condition follows this sequence (not all steps apply to the single-agent baseline):
\begin{enumerate}
\item \textbf{Initial proposals.} Each role-specialized agent calls \texttt{propose} once (proposal + rationale + confidence).
\item \textbf{Critique round.} Each agent calls \texttt{critique} on the full blinded proposal set (same labels for everyone; authorship hidden).
\item \textbf{Initial vote (diagnostic).} Each agent votes on the \emph{initial} proposals; ballots are logged as \texttt{ballots\_initial} and used for coordination metrics (vote shift, flip rate). They do \emph{not} select the delivered action by themselves.
\item \textbf{Revision.} Each agent calls \texttt{revise} using its own prior proposal, anonymized peer proposals, and the critique texts from round 1.
\item \textbf{Final vote.} Each agent votes on the \emph{revised} proposals; aggregation selects the winning tutoring move unless the protocol yields a tie or ambiguity handled below.
\item \textbf{Tie handling (if needed).} One extra critique-and-vote round on the tied candidates only; if still unresolved, deterministic role-priority fallback (\Secref{sec:tie}).
\end{enumerate}

\subsection{Outcome definitions and subsets}

\textbf{Simulated Performance Gains} for tables that report Initial / Final / Gain is $\text{Final}-\text{Initial}$, where both are the judge's continuous score in $[0,1]$ (\texttt{Judge.score} in \texttt{mas\_ed/simulation/judge.py}). For HumanEval, Initial/Final refer to the same student transcript (first vs.\ last attempt in the interaction), including any emitted code block.

\textbf{Task success} (\texttt{r.success} in logs): SciQ requires \texttt{final\_correctness}$\geq 0.75$. HumanEval additionally requires extracted student code to pass canonical unit tests (\texttt{code\_passed}). Heatmaps in \Secref{sec:persona} use these definitions per benchmark.

\textbf{Tutor-fired subset:} Judge correctness tables in \Secref{sec:out} restrict to interactions with $\texttt{turns\_taken}>0$ (tutor actually intervened). HumanEval code-pass statistics in our tables use the same restriction when labeled tutor-fired; aggregate metrics in \texttt{mas\_ed/evaluation/metrics.py} pool \emph{all} HumanEval rows unless a slice is applied downstream.

\textbf{Code execution:} Final and initial code pass use \texttt{mas\_ed/simulation/code\_judge.py} (extract Python from the student message, run canonical tests).

\subsection{LLM judge (prose scoring rubric)}

The judge is prompt-formatted to output \texttt{SCORE: <float>} and a one-line rationale. Its system prompt fixes anchor levels: $1.0$ fully correct and well explained; $0.7$ essentially correct with minor issues; $0.4$ partial understanding; $0.1$ mostly incorrect; $0.0$ no credit. For coding tasks, instructions tell the judge to score natural-language understanding against whether the described algorithm would match the canonical behavior, not to match source code verbatim.

\end{document}